%% file: main.tex
\documentclass[10pt, conference]{IEEEtran} 

\usepackage[T1]{fontenc}
\usepackage[utf8]{inputenc}
\usepackage{microtype} 

\usepackage{amsmath,amssymb,mathtools}
\usepackage{graphicx}
\usepackage{booktabs}
\usepackage{siunitx}
\usepackage{comment}
\usepackage{tabularx}
\usepackage{array}

\usepackage{cite} 
\usepackage{xcolor}
\usepackage[normalem]{ulem}
\usepackage{enumitem}
\setlist{itemsep=0.25em,topsep=0.3em,leftmargin=1.2em}

\usepackage[colorlinks=true, citecolor=blue, linkcolor=blue, urlcolor=blue]{hyperref}
\usepackage{orcidlink}
\usepackage[strings]{underscore}

\usepackage{listings}
\lstdefinestyle{sqdsty}{
  basicstyle=\ttfamily\small,
  columns=fullflexible,
  breaklines=true,
  frame=single,
  rulecolor=\color{black},
  numbersep=6pt,
  showstringspaces=false,
  tabsize=2
}
\title{SQD-Agent: LLM-driven agentic framework for Quantum Chemistry workflows}

\author{
\IEEEauthorblockN{Kislaya Tiwari\orcidlink{0009-0005-3287-273X}}
\IEEEauthorblockA{\textit{Indian Institute of Technology Delhi}\\
kislayatiwari@gmail.com}
\and
\IEEEauthorblockN{Anupama Ray\orcidlink{0000-0000-0000-0000}}
\IEEEauthorblockA{\textit{IBM Quantum, IBM Research}\\
anupamar@in.ibm.com}
}

\begin{document}

\maketitle

\begin{abstract}

Advances in quantum hardware and algorithms are positioning quantum-centric supercomputing as a promising paradigm for scientific applications. However, translating domain-specific problems into executable hybrid quantum-classical workflows remains a significant barrier for application researchers due to the required expertise in quantum algorithms, nuances in quantum programming, and hardware-aware system integration. At the same time, artificial intelligence has evolved into a transformative tool for scientific computing. Large language models (LLMs) are increasingly capable of interpreting natural-language intent, reasoning over complex workflows, and translating high-level objectives into executable code and building computational pipelines.
In this work, we introduce SQD Agent, an LLM-based agentic framework that translates natural-language user intent into executable workflows for Quantum Chemistry applications where algorithms from the Sample-Based Quantum Diagonalization (SQD) family are used. By automating this translation, SQD Agent reduces the level of human expertise and configuration overhead required, thereby simplifying experimentation in hybrid quantum-classical settings for application researchers new to quantum.
SQD Agent adopts a modular and extensible architecture that supports seamless integration of heterogeneous quantum backends, classical solvers, and workflow components, ensuring adaptability to rapidly evolving quantum ecosystems. The framework further incorporates interactive capabilities for on-demand profiling, bottleneck analysis, resource optimization, intelligent result caching, and convergence visualization. Key features include natural-language-driven quantum chemistry experiments, support for error mitigation on real quantum hardware, together with analysis of candidate mitigation schemes in terms of their potential error-recovery behavior and computational budget, helping users understand their practical trade-offs and decide which strategies to explore in subsequent experiments. This agentic system focuses on these key features, as these are common painpoints across domain experts, and such systems can advance quantum applications to a useful scale.

\end{abstract}


\input{Section_I/Introduction_v1}

\section{Architecture and Workflow}

\input{Section_II/Algo_and_Agent}

\input{Section_III/Execution_CLI_Examples}
\begin{figure}[t]
\centering
\includegraphics[width=\linewidth, height=10cm, keepaspectratio]{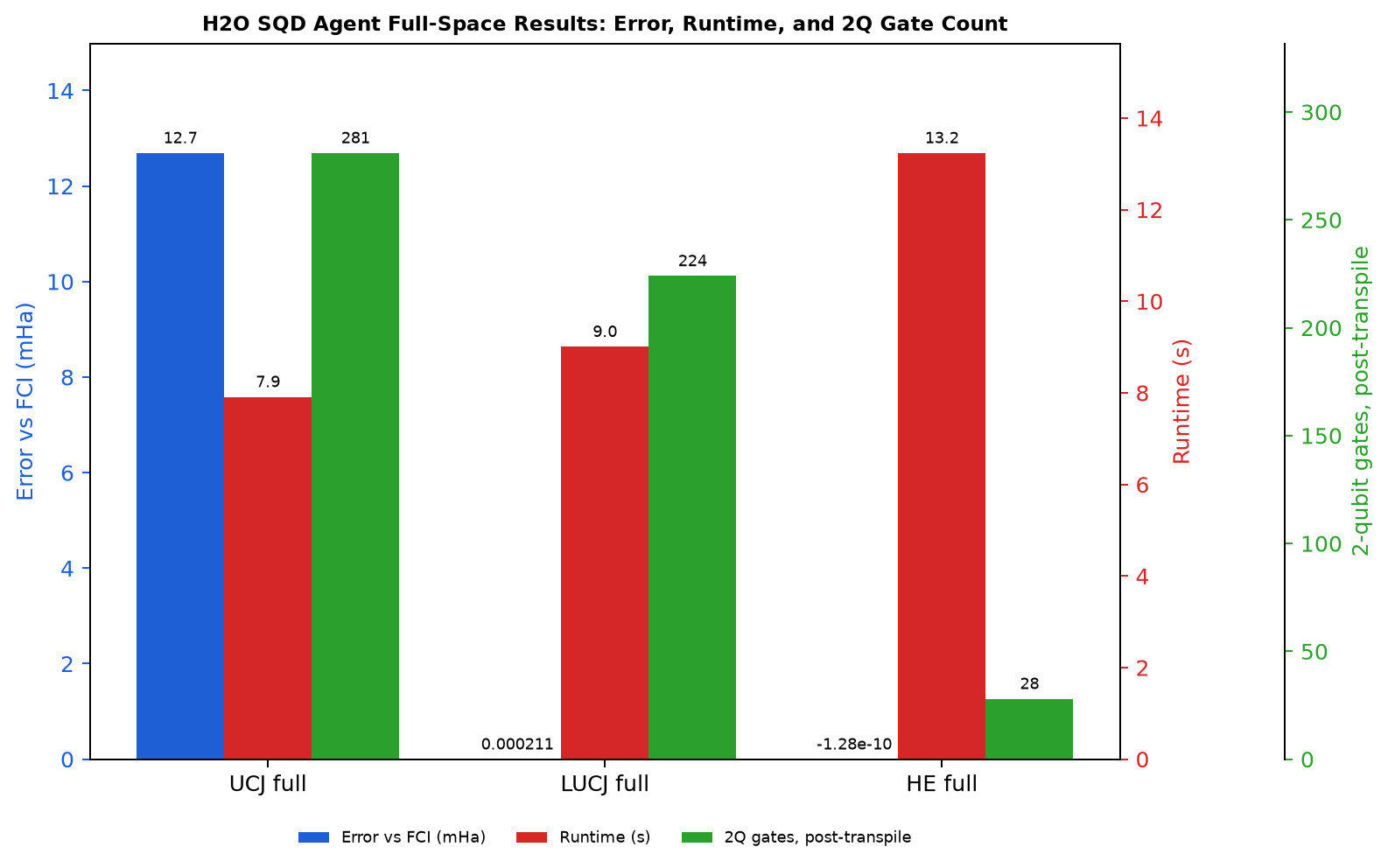}
\caption{Benchmarking ansatz across accuracy, runtime and two-qubit gate counts for $H_2O$ molecule}
\label{fig:sqd}
\end{figure}
\begin{figure}[t]
\centering
\includegraphics[width=\linewidth, height=10cm, keepaspectratio]{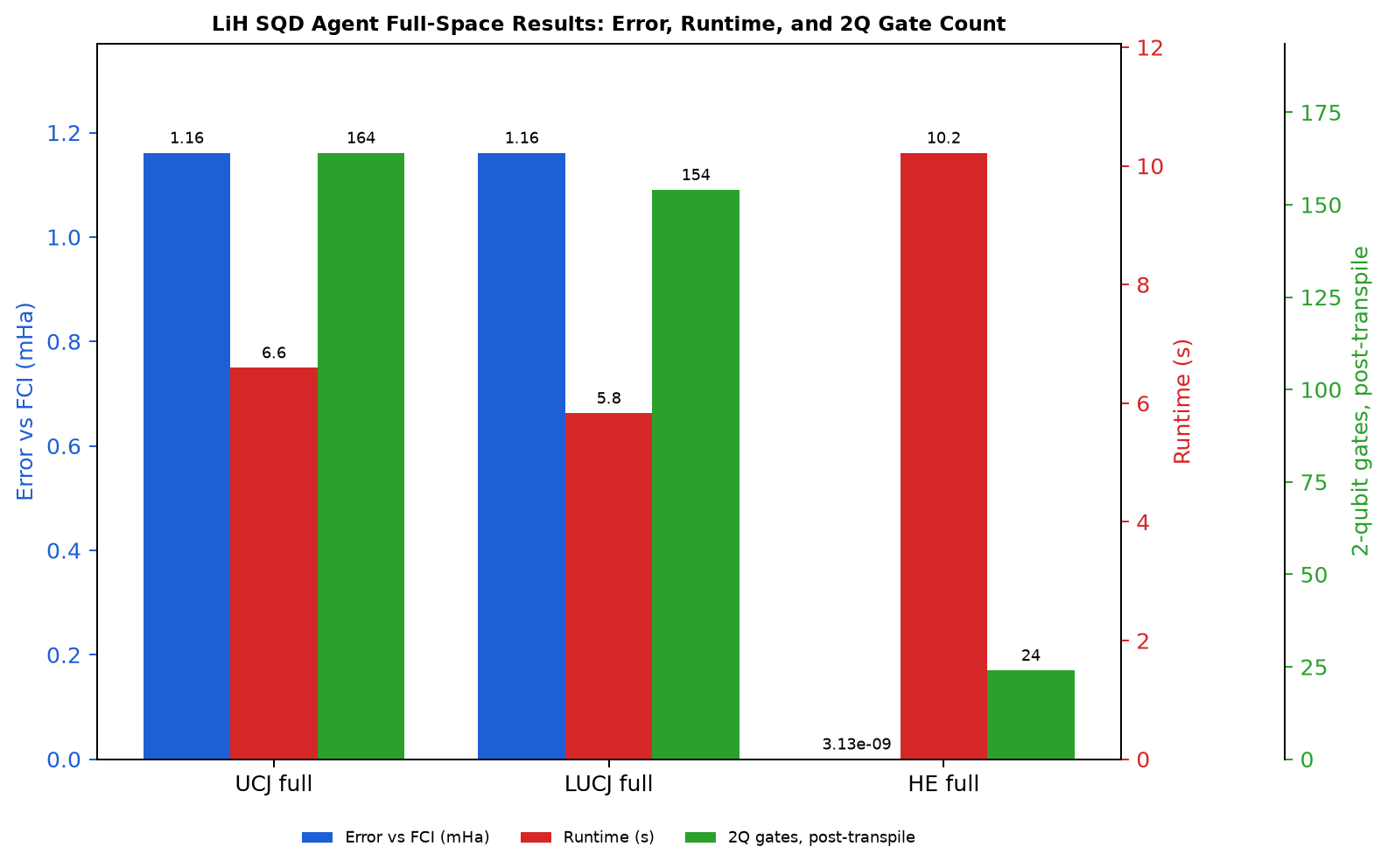}
\caption{Benchmarking ansatz across accuracy, runtime and two-qubit gate counts for $LiH$ molecule}
\label{fig:sim_h2o}
\end{figure}
\begin{figure}[t]
\centering
\includegraphics[width=\linewidth, height=10cm, keepaspectratio]{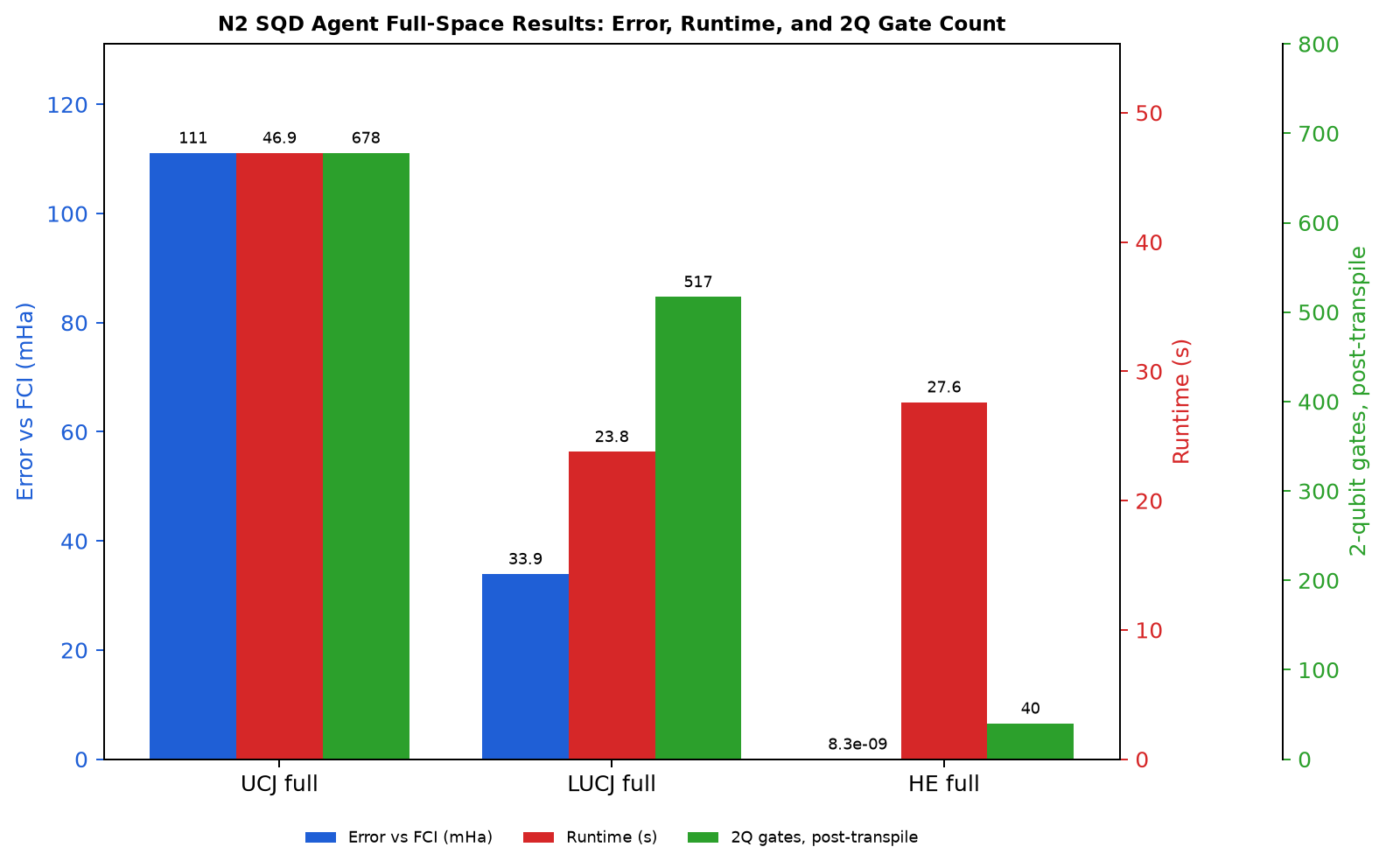}
\caption{Benchmarking ansatz across accuracy, runtime and two-qubit gate counts for $N_2$ molecule}
\label{fig:sim_lih}
\end{figure}

\begin{figure}[t]
\centering
\includegraphics[width=\linewidth, height=10cm, keepaspectratio]{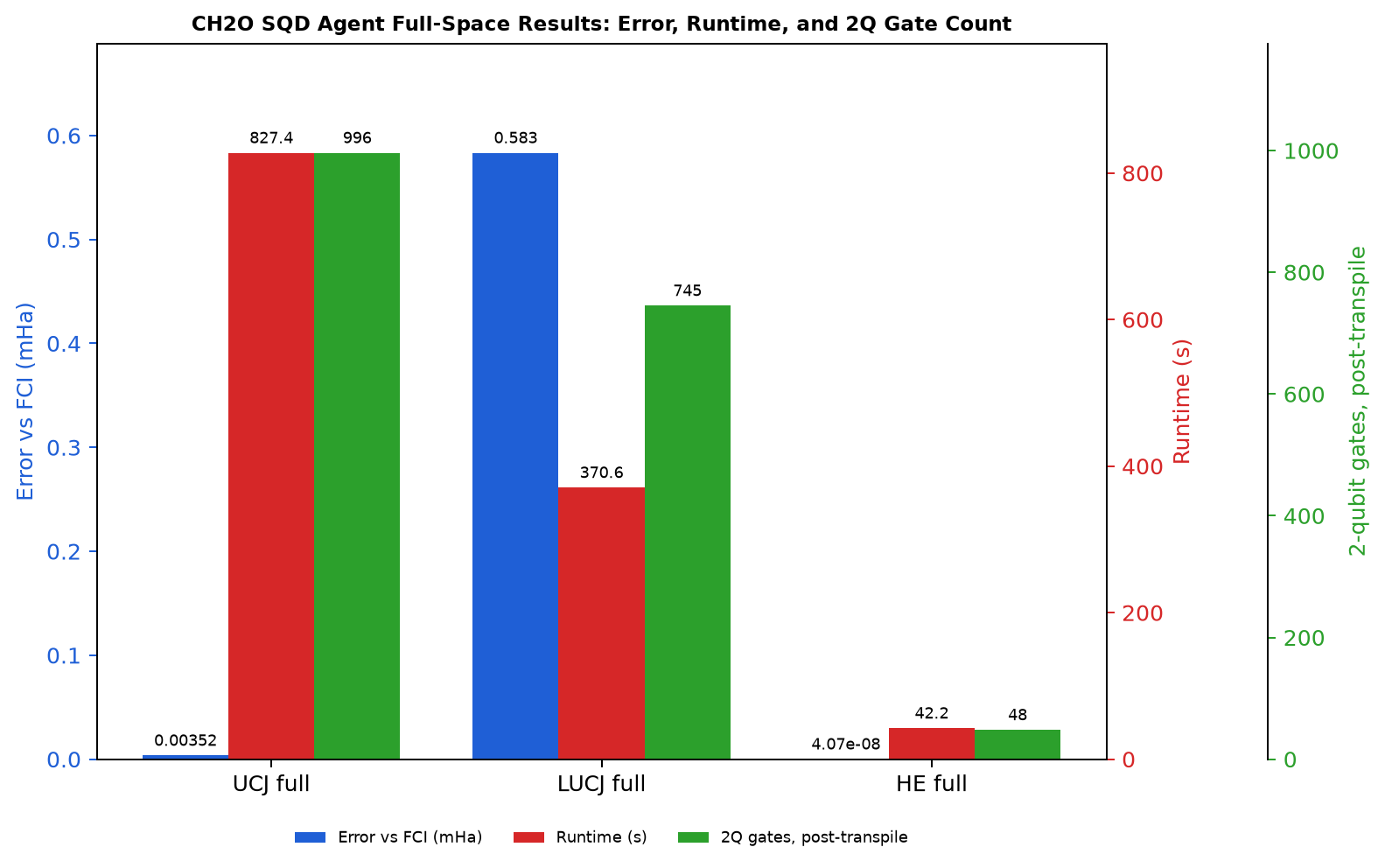}
\caption{Benchmarking ansatz across accuracy, runtime and two-qubit gate counts for $CH_2O$ molecule}
\label{fig:sim_ch2o}
\end{figure}
\begin{figure}[t]
\centering
\vspace{-2mm}
\includegraphics[width=\linewidth, height=10cm, keepaspectratio]{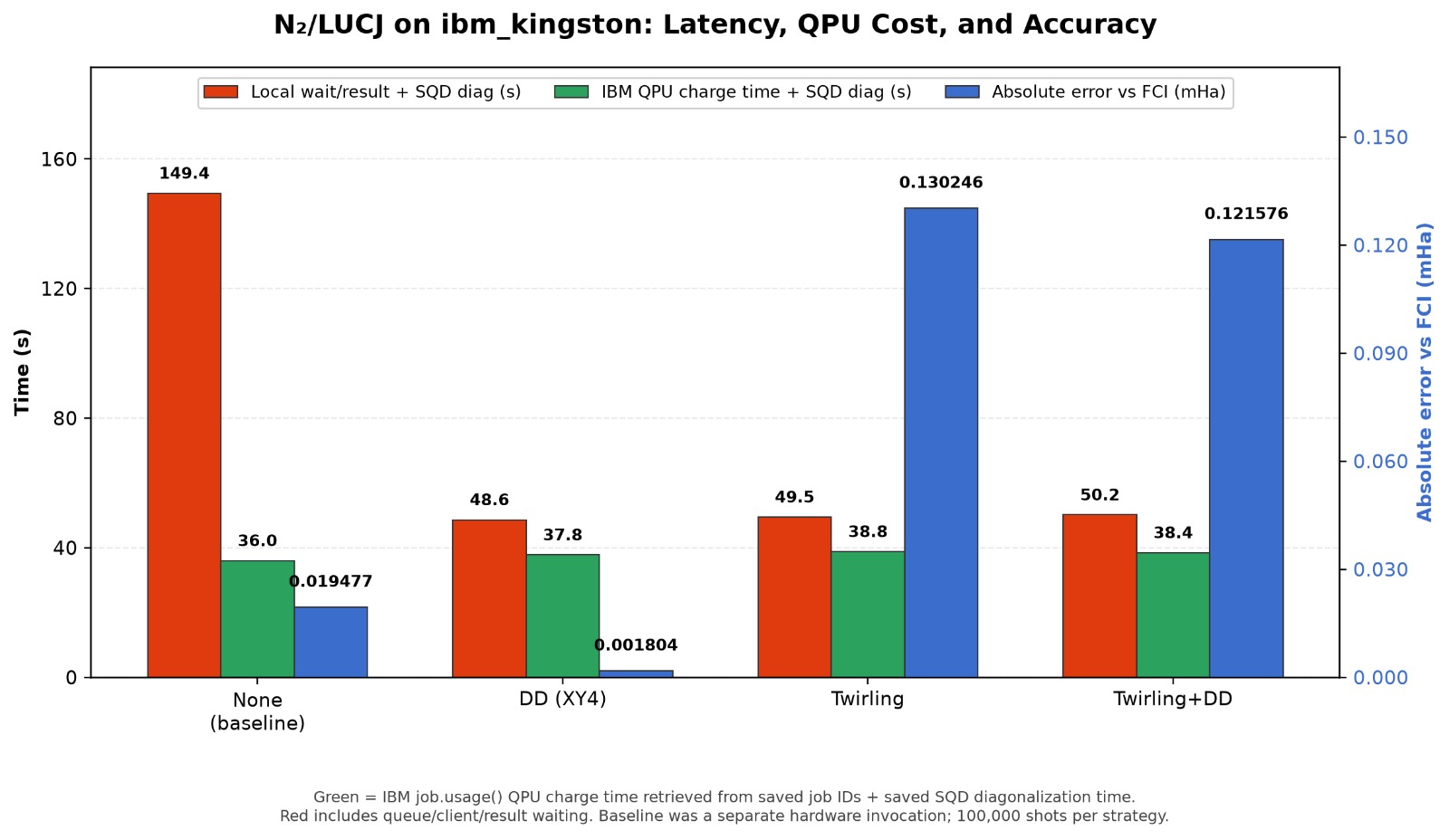}
\caption{Benchmarking Error mitigation across accuracy, runtime for $N_2$ energy estimation with LUCJ ansatz}
\label{fig:hw_n2}
\end{figure}
\input{conclusion/merged}


\section*{Data Availability}
All source code, demonstrations, simulators, and quantum hardware results supporting the findings of this study are maintained in a GitHub repository, whose link will be added after acceptance. The repository will be made publicly accessible upon publication.

\section*{Acknowledgements}
 OpenCode~\cite{opencode2026} with Gemini 2.5 Flash~\cite{gemini} was used for AI-assisted code/command support while processing and testing the \texttt{agent.md} workflow files. All AI-assisted text, code, commands, and outputs were constantly reviewed and verified by the authors.

\bibliographystyle{IEEEtran}
\bibliography{references}


\input{Appendix/Appendix_v2}

\end{document}

%% file: Section_I/Introduction_v1.tex
\section{Introduction}
\label{sec:Introduction}
\begin{figure*}[t]
\centering
\includegraphics[width=\linewidth, height=10cm, keepaspectratio]{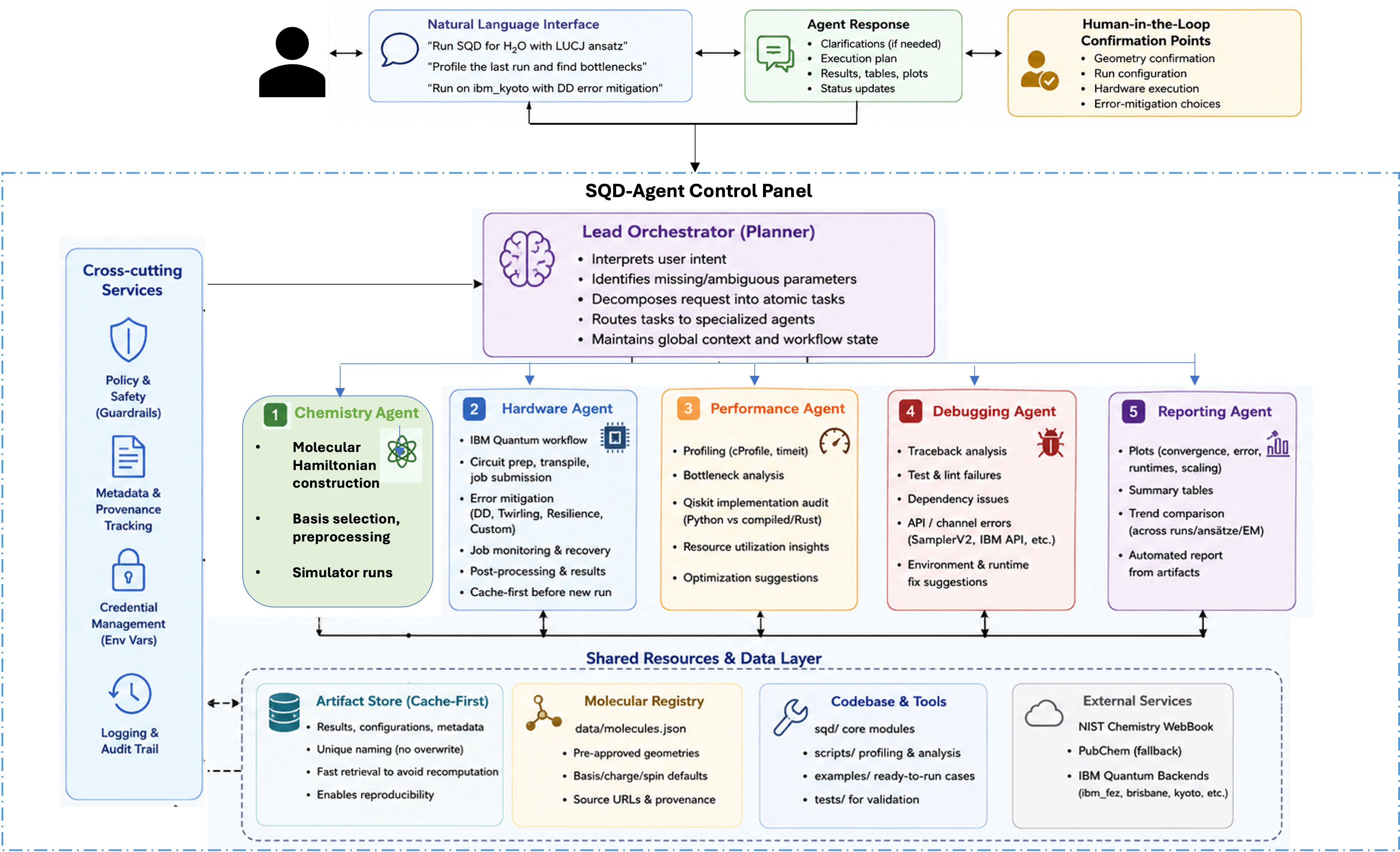}
\caption{Schematic representation of the SQD Agent workflow. A natural-language user request is interpreted by the lead orchestrator, which identifies the task, resolves missing parameters, and routes the request to specialized agents for chemistry setup, SQD-family execution, hardware submission, mitigation comparison, profiling, debugging, and reporting. The agents interact with a shared resource and data layer that stores molecular inputs, tool interfaces, cached artifacts, logs, and external-service connections. User supervision, guided by domain-specific knowledge, is maintained throughout the process.}
\label{fig:sqd_workflow}
\end{figure*}

Recent advances in hybrid quantum--classical algorithms, together with their use of heterogeneous computational platforms spanning quantum processors and classical CPU/GPU/HPC resources, have enabled quantum chemistry studies of molecular systems that were previously difficult to simulate using purely classical or purely quantum approaches. Computational chemistry is central to applications such as catalyst design, materials discovery, and drug development; however, accurate electronic-structure calculations remain challenging because the dimension of the many-electron Hilbert space grows rapidly with the number of active orbitals. Classical methods such as coupled cluster, complete active-space approaches, and selected configuration interaction provide powerful approximations, yet their computational cost and configuration complexity motivate hybrid quantum-classical approaches for near-term devices.

Sample-based subspace iteration methods have emerged as a promising family of algorithms for quantum chemistry on noisy quantum hardware through quantum-centric supercomputing platforms. Sampling-based quantum diagonalization (SQD) uses a quantum circuit to generate measurement samples, constructs a selected configuration subspace from those samples, and performs classical diagonalization of the projected Hamiltonian. This structure makes SQD attractive for near-term systems because the quantum processor is used primarily as a sampler, while the expensive eigenvalue problem is delegated to classical resources. Recent work has demonstrated the potential of SQD-style workflows for molecular systems beyond the scale of exact diagonalization~\cite{robledo2024beyond} and even for protein complexes spanning 12,635 atoms~\cite{merz2026crossing12000atombarrierheterogeneous}. This progress has further motivated several extensions of subspace-iteration-based SQD, including Krylov-enhanced, randomized, and randomized SQD variants~\cite{Yu2025SKQD,Piccinelli2025SqDRIFT,Graves2026RESD}.

Despite this algorithmic progress, executing computational chemistry experiments remains difficult for many application researchers because of the steep learning curve. Users must possess expert knowledge of computational chemistry protocols, quantum algorithms, hardware and software usage, and coding across the full stack. A complete workflow requires selecting a molecular geometry and basis set, choosing an active space, identifying a suitable algorithm, constructing the corresponding quantum circuits, benchmarking on simulators when possible, efficiently transpiling circuits for execution on quantum hardware, and performing classical diagonalization and configuration recovery on classical infrastructure. Furthermore, when executing on quantum devices, users must understand shot budgets, apply and compare different error-mitigation strategies, interpret convergence behavior, and explain the resulting metrics. These steps require expertise across quantum chemistry, quantum programming, hardware execution, and classical post-processing. Existing software libraries such as PySCF, Qiskit, \texttt{ffsim}, and \texttt{qiskit-addon-sqd} provide the necessary computational building blocks, but users must still manually assemble them into a reproducible and hardware-aware workflow~\cite{sun2018pyscf,qiskitnature2022}.

With the recent success of LLMs and agentic systems, several research efforts have explored their use in computational chemistry, quantum computing, and scientific discovery, including ChemCrow~\cite{ref50}, Coscientist~\cite{ref51}, SciAgents~\cite{li2023sciagents}, ChemAgent for memory-enhanced chemical reasoning~\cite{tang2025chemagent}, tool-augmented ChemAgent for chemistry problem solving~\cite{yu2025tooling}, El Agente Q~\cite{zou2025elagenteq}, autonomous LLM agents for quantum-physics research~\cite{arlt2025autonomous}, and broader AI-driven approaches to quantum computing~\cite{alexeev2025artificial}. Most of these systems aim to support scientific literature understanding and the investigation of new scientific hypotheses through planning, tool calling, and coordination across different software tools and databases. However, they primarily target classical chemistry workflows or general scientific tool use, and do not directly address the quantum-native requirements of quantum computational chemistry workflows.

Two recent works focus on quantum chemistry: El Agente Quntur \cite{PerezSanchez2026AgenteQunturChem} and El Agente Cuantico \cite{Gustin_2026}. El Agente Quntur addresses high-level chemical queries using ORCA 6.0 and performs in-depth research across software documentation and scientific literature. In terms of methodology, El Agente Cuantico is most similar to SQD-Agent, as it employs a multi-agent AI system to automate quantum simulation workflows across different quantum software frameworks. In contrast, SQD-Agent specifically targets quantum chemistry computations supported by current quantum algorithms and integrates with real quantum devices for execution. SQD-Agent also focuses on intelligently selecting error mitigation strategies to improve result quality on noisy quantum hardware while educating users about the underlying techniques. Furthermore, it adopts a human-in-the-loop approach that incorporates resource profiling and management across classical and quantum computing resources to support quantum-centric supercomputing workflows.

has optional human in loop (allows human input) and removed hard coded policies to enable scientific reasoning across tools via letting the LLM read api documentation thus it can update itself as the software updates and manually nothing has to be done.

In this paper, we present a multi-agent framework for quantum chemistry experiments that leverages LLMs for decision making and tool calling across the software stack, together with specialized quantum agents that manage sampling-based subspace iteration methods and other quantum--classical research steps. We call this the \textbf{SQD Agent}, since SQD was the first sampling-based subspace iteration method to demonstrate quantum-chemistry calculations beyond the scale of exact diagonalization. This LLM-based agentic framework orchestrates quantum-chemistry workflows on simulators and near-term quantum hardware for different versions of SQD and related subspace-iteration methods. The framework separates language-level planning from deterministic execution and is explained in detail in Figure \ref{fig:sqd_workflow}. A lead orchestrator interprets the user's natural-language request and dispatches it to the appropriate specialized agent. At present, the framework includes specialized agents for SQD-family execution on simulators, hardware execution, mitigation comparison, profiling, debugging, and reporting. These agents construct validated task-specific specifications that are passed to a modular execution layer built on PySCF, Qiskit, \texttt{ffsim}, and \texttt{qiskit-addon-sqd}. The execution layer then performs electronic-structure preparation, circuit construction, sampling, SQD post-processing, subspace diagonalization, artifact storage, and report generation.


The main contributions of this paper are as follows:
\begin{itemize}
\item We present a multi-agent system for advancing scientific discovery in quantum chemistry by agent-driven experiment orchestration, thereby accelerating experimentation across diverse levels of user expertise through natural-language inputs.
\item The framework is implemented in a modular manner, enabling the integration of additional algorithms (currently from the SQD family such as extended SQD~\cite{barison2025extendedsqd}, and SQdrift~\cite{Piccinelli2025SqDRIFT}) as well as tools (such as Fulqrum~\cite{nation2026fulqrum, fulqrum2026}) and other tools for profiling, logging, benchmarking, etc. 
\item We develop a hardware agent that supports quantum-native circuit optimization, quantum execution, job retrieval, mitigation-strategy suggestion, and post-run comparison of observed accuracy-cost trade-offs, where cost includes shots, QPU jobs, runtime, and classical post-processing.
\item We evaluate the framework through experiments on six molecular systems using both simulators and IBM quantum systems, demonstrating end-to-end agent-mediated workflow execution and detailed result dashboards.

\end{itemize}

Rather than coordinating only classical chemistry packages as in as El Agente Q \cite{zou2025elagenteq}, SQD-Agent targets a quantum-native workflow in which the agent manages quantum circuit generation, transpiling strategies for quantum hardware, efficient sampling from quantum devices, various error mitigation policies while optimizing recovery and budget, SQD diagonalization, and performance-aware reporting. The framework is therefore positioned as an agentic execution and analysis layer for near-term quantum-centric chemistry experiments, and not as a replacement for the underlying SQD family of algorithms.

%% file: Section_II/Algo_and_Agent.tex
\subsection{Agent-Callable SQD Execution Engine}
\label{sec:sqd_execution_layer}

The algorithmic core of SQD Agent is implemented as a modular Python execution engine with an agent-facing control interface. The execution routines can be invoked either by the mid-level SQD agent or directly through the command-line interface, enabling the system to support both conversational use and reproducible scripted execution. The role of the core execution engine is to receive a validated workflow request, execute an SQD-family quantum-chemistry calculation, and return numerical outputs, diagnostic traces, and metadata for reproducibility. The numerical outputs and diagnostics support energy analysis, convergence inspection, resource comparison, profiling, and hardware-aware interpretation, while the metadata preserves the configuration needed to reproduce or retrieve a run. This design separates language-level planning from deterministic numerical execution, that is, the LLM-based agent chooses, validates, and explains the workflow configuration, while electronic-structure calculations, circuit construction, quantum sampling, subspace diagonalization, and post-processing are performed by fixed library-backed routines. Since the engine is organized through modular interfaces for chemistry preparation, ansätz construction, backend execution, subspace solving, mitigation, profiling, and reporting; new features and parameters can be added as improved libraries, solvers, backend services, and SQD-family components become available.

To make the separation between agent planning and operational execution, the SQD-execution specialist communicates with the execution layer through a structured run contract, denoted here as \texttt{SQDRunSpec}. This contract is specific to SQD execution tasks and does not represent the full reasoning capacity of the agentic framework. It records the information needed to perform a reproducible SQD run, including the molecular geometry, basis set, charge and spin, active-space policy, ansätz family, backend target, shot count, SQD iteration count, batching mode, transpilation profile, error-mitigation policy, profiling flag, and cache policy. The execution layer returns an execution-specific \texttt{SQDRunResult} containing the total energy, available classical reference energies, circuit metrics, convergence trace, timing breakdown, artifact paths, and, for hardware runs, a mitigation report. Other specialists, such as the performance, hardware, debugging, and reporting agents, may consume or extend these artifacts according to their own task-specific policies. By requiring all agent decisions to pass through this structured contract, the framework prevents free-form code execution and makes every run auditable.

\paragraph{Electronic-structure and active-space preparation}
The workflow begins with electronic-structure preparation using PySCF. Given a molecular specification, the chemistry layer constructs a Hartree--Fock reference and forms the one- and two-electron integrals in the molecular-orbital basis together with the core energy. When requested, MP2, CCSD, CASCI, or FCI reference values are computed when feasible and stored for later comparison. For reduced active-space calculations, the active-space module selects a chemically relevant orbital window using one of the implemented policies: a HOMO--LUMO window, a natural-occupation criterion, or a CCSD-amplitude-based orbital score. A capacity-preserving refit ensures that the selected active space can accommodate the active electrons, avoiding silent electron loss during truncation.

The ansätz layer constructs the quantum state-preparation circuit under a Jordan--Wigner mapping. The current implementation supports a Hartree--Fock reference circuit, a Unitary Cluster Jastrow (UCJ) circuit initialized from CCSD amplitudes through \texttt{ffsim}, a localized UCJ proxy that restricts excitation locality to reduce two-qubit gate count, and a fixed hardware-efficient circuit. The ansätz interface is modular: additional ansätz families, initialization rules, locality constraints, or hardware-adapted circuit templates can be added as new construction modules and exposed to the user through the agent interface. Thus, a user may request an existing ansätz, compare multiple ansätze, or, when supported by the codebase, invoke newly added ansätz variants through the same natural-language workflow. By default, each constructed circuit is saved with ansätz-related metadata such as logical qubit count, pre- and post-transpilation depth, two-qubit gate count, and total gate count. Additional ansätz diagnostics, circuit summaries, or comparative plots can also be generated by the agent on demand from the stored artifacts.

\paragraph{SQD execution loop}
The SQD loop follows the sample-based diagonalization workflow: the prepared circuit is sampled on the selected backend, measured bitstrings are filtered or recovered to enforce the target electron number, configurations are batched into selected subspaces, and the molecular Hamiltonian is projected and diagonalized in those subspaces. The implementation reuses the \texttt{qiskit-addon-sqd} primitives for configuration recovery and projected diagonalization rather than reimplementing the SQD solver itself. The configuration pool can be handled in single-batch, merged-batch, or independent-batch mode. Merged batches are deduplicated and pooled before diagonalization, increasing the subspace dimension and may improve accuracy, while independent batches reduce the size of each diagonalization and expose parallelism. The per-iteration best energy, subspace size, and occupancy information are recorded as convergence diagnostics.  

The current diagonalization path delegates the selected-configuration problem to \texttt{solve_fermion}, which constructs and solves the projected Hamiltonian using sparse classical linear algebra. However, this solver is not an architectural constraint. The execution layer treats the diagonalizer as a replaceable solver component: future selected-CI solvers, GPU-accelerated eigensolvers, distributed sparse diagonalizers, or SQD-family variants can be added behind the same interface, provided they consume the selected configurations and molecular integrals and return the projected ground-state estimate and associated diagnostics.

\paragraph{Backend and hardware abstraction}
The execution engine separates SQD logic from backend-specific execution. Simulator runs are currently supported through Qiskit Aer backends such as statevector, matrix-product-state, and GPU-enabled simulation when available. Hardware runs use Qiskit Runtime and IBM Quantum backends, with transpilation profiles controlling the trade-off between compilation time and circuit quality. After sampling, both simulator and hardware results are converted into the same bitstring-count interface, allowing the SQD post-processing loop to remain backend independent.

This backend abstraction is important for extensibility. Support to IBM Quantum devices is the current hardware path, but the architecture does not assume IBM-specific hardware at the SQD level. Additional quantum processors or cloud providers can be integrated through backend adapters that implement the same operations: circuit submission, job monitoring, result retrieval, metadata recording, and conversion of measurement outcomes into the common sampling format. Thus, new backends can be added without changing the agent-orchestration logic or the SQD diagonalization pipeline.

\paragraph{Error Mitigation for hardware execution}
For physical-device runs, the execution layer exposes error mitigation as a configurable policy object rather than as a hard-coded option. If the user does not specify a mitigation strategy, or explicitly asks for guidance, the hardware agent can suggest compatible strategies before execution using the backend capabilities, circuit depth, two-qubit gate count, shot count, and user objective. The objective may be to minimize runtime, improve robustness, or obtain a comparative mitigation study. The agent may then run an unmitigated baseline, execute one selected mitigation strategy, or sweep over multiple supported strategies.

When multiple strategies are executed, the system records for each case the backend, transpilation profile, shot count, QPU time, classical diagonalization time, recovered SQD energy, and deviation from the selected reference where available. These entries are collected into a \texttt{MitigationReport}, which compares the observed time--accuracy trade-off across mitigation choices. The recommendation is therefore context-dependent: before execution it is based on compatibility and expected overhead, while after execution it is based on the measured behavior for the present molecule, ansätz, backend, and shot budget. The framework does not assume that one mitigation method is universally optimal across all SQD workloads.

\paragraph{Resource logging, profiling, and artifact management}
The execution layer records the resource information needed to interpret each SQD run. On the quantum side, the saved metadata includes the number of logical qubits, transpiled circuit depth, two-qubit gate count, total gate count, shot count, backend name, and QPU runtime when available. On the classical side, the framework records the wall-clock time of major stages such as circuit sampling, SQD post-processing, subspace diagonalization, and result serialization. These measurements are not intended to replace a full low-level performance model; rather, they provide a lightweight, reproducible timing breakdown that allows the user to perform further analyses on the fly.

 When requested, the agent can invoke an optional profiling mode that generates a more detailed timing report for the same workflow. The current design therefore supports two levels of performance analysis: default resource logging for every saved run, and on-demand profiling for deeper inspection before optimization or demonstration experiments. The resulting timing summaries, convergence traces, circuit metrics, and run metadata are stored with the output artifacts so that the reporting agent can later generate tables, plots, and qualitative bottleneck summaries without rerunning the calculation.
 
All data-producing actions write timestamped artifacts under the benchmark archive. Before launching a new computation, the agent constructs a run signature from the molecule, basis, charge, spin, ansätz, active-space policy, backend, shot count, SQD settings, mitigation policy, random seed when available, and software environment. If a compatible result already exists, the cache-first policy reuses the stored artifact rather than recomputing it. For hardware jobs, saved job identifiers allow interrupted runs to be retrieved after completion. The same artifact store supplies convergence plots, benchmark tables, mitigation summaries, and reproducibility metadata to the reporting agent.

Overall, the execution layer provides the deterministic substrate on which the agentic framework operates. It combines established quantum-chemistry and quantum-computing libraries with structured workflow control, backend abstraction, mitigation suggestion, profiling, and artifact management. This makes SQD Agent extensible to new SQD-family algorithms, new classical solvers, and new quantum hardware backends while preserving the reproducibility and safety constraints required for near-term quantum-centric experiments.

\subsection{Agentic Orchestration and Workflow Control}

The automation layer connects the deterministic SQD execution engine of Section~II-B to a natural-language interface. Rather than allowing the language model to directly generate and execute arbitrary code, SQD Agent uses the LLM as a workflow planner that produces a validated, task-dependent execution specification. A user request is first interpreted by a lead orchestrator, which identifies the intended task---for example simulator execution, hardware execution, profiling, debugging, mitigation comparison, or report generation---and dispatches the request to the appropriate specialist agent. The specialist then constructs the corresponding structured request. For SQD energy-estimation tasks, this request is represented as an \texttt{SQDRunSpec}, whose fields are populated according to the task requested by the user. For example, a simulator benchmark may require molecule, basis, ansätz family, active-space policy, shot count, batching mode, and profiling options, whereas a hardware run additionally requires backend selection, transpilation profile, job policy, and error-mitigation settings. The execution layer accepts only such validated structured requests and returns task-specific artifacts, such as an \texttt{SQDRunResult} containing the estimated energy, reference energies where available, circuit metrics, convergence trace, timing breakdown, and artifact paths, or a hardware-specific mitigation report when mitigation strategies are executed or compared.

This separation between language-level planning and deterministic execution is central to the reliability of the framework. The lead orchestrator does not perform quantum-chemistry computation; it routes tasks, resolves missing information, and dispatches the request to the appropriate specialist. The execution specialist handles simulator-based energy estimation and multi-ansätz benchmarking. The hardware specialist manages backend selection, transpilation, job submission, job retrieval, mitigation strategy selection, and post-processing of hardware results. The performance specialist invokes profiling and bottleneck-analysis routines, while the reporting specialist generates tables, convergence plots, and accuracy--cost summaries from stored artifacts. Here, cost refers to practical execution resources such as shot count, number of hardware jobs, QPU runtime, queue or retrieval overhead, classical post-processing time, and diagonalization time. Each specialist is restricted to a narrow operational scope, which makes the workflow auditable and prevents a single agent from silently changing scientific or hardware parameters.

The framework uses human-in-the-loop validation for choices that affect scientific integrity or quantum-resource consumption. This validation is designed to rely on the user's domain expertise rather than on low-level technical knowledge of quantum software or hardware execution. For instance, if the user provides a complete molecular geometry, it is used directly. If only a molecular name is supplied, the agent first searches the local molecule registry and otherwise retrieves a candidate structure from an external chemistry source. The selected geometry is then shown to the user for confirmation before the run proceeds, allowing the user to judge whether the structure is scientifically appropriate while the agent handles the technical conversion into executable input. Similarly, hardware execution requires explicit confirmation of the backend, ansätz, shot count, and mitigation strategy before a QPU job is submitted. This design keeps scientific oversight with the researcher while delegating command construction, software configuration, data movement, hardware interaction, and post-processing to the agent.

SQD Agent also follows a cache-first execution policy. Before launching a new computation or analysis task, the specialist agent constructs a task-dependent signature from the parameters relevant to that request. For an SQD-family execution task, this signature includes the molecule, basis, charge, spin, ansätz, active-space selection, backend, shot count, SQD settings, mitigation policy, random seed when available, and software environment. For profiling and bottleneck analysis, the signature additionally records system-dependent information such as processor and memory configuration, available GPU support, operating system, Python environment, library versions, and linear-algebra thread settings, since performance measurements are meaningful only relative to the machine on which they were obtained. For reporting tasks, the specification instead points to stored result artifacts, convergence traces, circuit metrics, and mitigation tables from which plots or summaries should be generated. If a compatible artifact already exists for the requested task, the result is summarized and reused rather than recomputed. For interrupted hardware tasks, stored job identifiers are used to retrieve completed results from the backend. This policy reduces redundant simulator and QPU cost while preserving traceability through timestamped metadata, saved commands, convergence traces, circuit metrics, profiling records, and generated reports.

Because the execution-layer extensibility is handled through the modular interfaces described in Section~\ref{sec:sqd_execution_layer}, the agent layer does not need to change when new ansätze, SQD-family variants, solvers, or hardware backends are added. The specialist agents only need to expose the new option through the corresponding task-specific specification and validation policy.

This distinguishes SQD Agent from general chemistry agents such as El Agente Q. While those systems demonstrate the usefulness of LLM agents for coordinating classical quantum-chemistry software, SQD Agent targets a quantum-native workflow in which the agent must manage quantum sampling, hardware-aware transpilation, shot budgets, mitigation policies, classical subspace diagonalization, and quantum--classical performance trade-offs. The contribution is therefore not a general-purpose chemistry chatbot, but an agentic control layer for reproducible and performance-aware SQD-family workflows on near-term quantum-centric systems which can be supervised by the user's domain-specific knowledge.

Table~\ref{tab:sqd_agent_layers} summarizes the workflow layers of SQD Agent. The table emphasizes the main distinction from general chemistry agents such as El Agente Q: SQD Agent is designed around quantum-native SQD execution, hardware submission, mitigation comparison, and quantum--classical performance reporting.

\begin{table}[t]
\centering
\caption{Workflow layers of SQD Agent and their distinction from general chemistry agents.}
\label{tab:sqd_agent_layers}
\scriptsize
\renewcommand{\arraystretch}{1.12}
\setlength{\extrarowheight}{4pt}
\begin{tabularx}{\columnwidth}{p{0.31\columnwidth} X}
\toprule
\textbf{Layer} & \textbf{Description} \\
\midrule

Lead orchestrator &
Interprets natural-language intent using the LLM, plans a workflow and routes the task to a specialist. \\

SQD-family algorithm execution agent &
Converts energy-estimation requests into validated SQD-family execution specifications, enabling simulator runs and ansätz comparisons without manual quantum-workflow configuration. \\

Hardware execution agent &
Manages backend selection, transpilation, QPU submission, job retrieval, and hardware-result processing, extending agentic chemistry workflows from classical software to real quantum devices. \\

Mitigation advisor &
Suggests compatible mitigation strategies before execution and compares observed accuracy--cost trade-offs after mitigation sweeps, where cost includes shots, QPU jobs, runtime, and post-processing time. \\

Performance agent &
Produces resource and timing summaries for quantum--classical SQD workflows, exposing the cost split between sampling, post-processing, and subspace diagonalization. \\

Reporting agent &
Generates convergence plots, benchmark tables, and mitigation summaries from stored artifacts, enabling SQD experiments to be studied across sessions. \\

Debugging agent &
Assists with dependency, API, command, and execution failures while preserving the validated run specification and artifact provenance. \\

SQD execution layer &
Provides the deterministic quantum-native substrate built on PySCF, Qiskit, \texttt{ffsim}, and \texttt{qiskit-addon-sqd}, separating SQD computation from LLM planning and remaining extensible to new solvers, ansätze, and backends. \\

\bottomrule
\end{tabularx}
\end{table}

%% file: Section_III/Execution_CLI_Examples.tex
\section{Experimental Evaluation of Agentic SQD Workflows}
\label{sec:experiments}

This section presents the experimental evaluation of SQD Agent as an end-to-end agentic workflow system, performed with OpenCode \cite{opencode2026}. The experiments are organized to demonstrate how the framework translates user intent into reproducible SQD-family executions, exposes quantum--classical resource costs, reuses stored artifacts, generates convergence and timing reports, and compares mitigation strategies on a real quantum backend. We structure the evaluation around five questions:
\begin{itemize}
\item \textbf{Q1:} Can a natural-language prompt be converted into a valid SQD-family workflow?
\item \textbf{Q2:} Do the generated workflows recover reference energies on simulator benchmarks?
\item \textbf{Q3:} Does the framework expose quantum and classical resource costs?
\item \textbf{Q4:} Can cached artifacts, convergence traces, and timing summaries support reproducible reporting?
\item \textbf{Q5:} Can the hardware agent execute and compare mitigation strategies on a real QPU?
\end{itemize}

\subsection{Evaluation Protocol and Setup}

The core software stack consists of PySCF for electronic-structure preparation, Qiskit and Qiskit Aer for circuit construction and simulation, \texttt{ffsim} for fermionic state preparation, and \texttt{qiskit-addon-sqd} for sample-based diagonalization. Hardware execution used Qiskit Runtime on the 156-qubit IBM Quantum backend \texttt{ibm_kingston}. For each saved run, SQD Agent records the task specification, command-line invocation or agent-generated task request, software environment, linear-algebra thread settings, circuit metrics, convergence trace, timing summary, and output artifacts. Hardware runs additionally store backend metadata, mitigation configuration, and job identifiers so that interrupted executions can be retrieved rather than resubmitted.

The simulator benchmarks use the STO-3G basis and compare the ansätz families available in the current implementation: Hartree--Fock (HF), Unitary Cluster Jastrow (UCJ), localized UCJ (LUCJ), and a fixed hardware-efficient ansätz (HE). Unless otherwise stated, simulator benchmarks use \num{300,000} shots per circuit, samples-per-batch=100,000, max\_iters=8, seeded MPS seed\_simulator=42. Reference energies are computed using classical methods when feasible, including RHF, MP2, CCSD, CASCI, and FCI. The hardware experiment is performed with the N\textsubscript{2}/LUCJ workflow to test backend submission, result retrieval, SQD post-processing, and mitigation comparison on a physical quantum processor.

\subsection{Natural-Language to SQD Workflow Validation}

To test whether user intent can be converted into an executable SQD-family workflow, we used the prompt:
\begin{quote}
\emph{``Run SQD energy estimation for formaldehyde in the STO-3G basis and compare ansätze.''}
\end{quote}
The lead orchestrator identified the task as simulator-based SQD benchmarking and dispatched it to the SQD-family algorithm execution agent. The specialist agent resolved the molecule, selected the STO-3G basis, generated a benchmark-style execution specification, and invoked the execution layer for the available ansätz families. The resulting artifacts included an ansätz-comparison table, circuit metrics, convergence traces, and saved  (see Table~\ref{tab:h2co_energy_time}). On request, the reporting agent used the same artifact directory to generate convergence plots and summary tables without requiring the user to manually reconstruct the command-line invocation.

This example demonstrates the intended division of labor. The user supplies the scientific objective in natural language, while the agent handles workflow construction, parameter resolution, execution, artifact storage, and report generation. The same run remains reproducible because the generated execution specification and artifacts are saved with the result.

\subsection{Simulator Accuracy and Resource Metrics}

Table~\ref{tab:sim_summary} summarizes the simulator benchmarks. The goal of these experiments is to validate that SQD Agent correctly orchestrates SQD-family workflows and records the accuracy and resource information needed for comparison. Across the tested molecules, the correlated UCJ/LUCJ-style ansätze recover reference energies with small deviations in the full-space calculations, while active-space runs expose the expected truncation error when the active window omits correlation-relevant orbitals.

\begin{table}[t]
\centering
\caption{Summary of SQD Agent benchmarks in STO-3G basis performed on qiskit quantum simulator. Detailed results are in Appendix~\ref{appendix:results}}
\label{tab:sim_summary}
\scriptsize
\renewcommand{\arraystretch}{1.12}
\setlength{\extrarowheight}{4pt}
\begin{tabular}{|c c c|}
\hline
\textbf{Molecule} & \textbf{Best SQD ansätz} & \textbf{Error vs. FCI}  \\
\hline
LiH &
UCJ/LUCJ &
$+1.161$~mHa  \\

H\textsubscript{2}O &
LUCJ &
$+0.000$~mHa  \\

N\textsubscript{2} &
UCJ &
$+0.418$~mHa \\

\hline
\end{tabular}
\end{table}

The recorded metrics include logical qubit count, transpiled depth, two-qubit gate count, total gate count, shot count, wall-clock runtime, and SQD convergence traces. These measurements allow the reporting agent to compare ansätze not only by final energy but also by execution cost. For example, UCJ and LUCJ provide chemically accurate or near-chemical-accuracy estimates in the tested full-space cases, while the hardware-efficient circuit has lower structural complexity but does not consistently provide the same accuracy across molecules. Such comparisons are useful for application researchers because they expose the practical trade-off between circuit cost and energy recovery.

\subsection{Profiling, Caching, and Convergence Reporting}

The execution layer described in Section~\ref{sec:sqd_execution_layer} records the resource and timing information needed to interpret each saved SQD run. A default set of metadata is stored for every execution, including the molecule, basis, ansätz, backend name, shot count, qubit count, total circuit depth, two-qubit gate count, total gate count, convergence trace, and total runtime. Additional details are recorded depending on the user request and task type. For example, a hardware run stores backend metadata, error mitigation configuration, QPU runtime, and job identifiers, while a profiling request records a more detailed timing breakdown for stages such as sampling, SQD post-processing, subspace diagonalization, and result serialization. These timing reports are interpreted relative to the recorded system configuration, since profiling can depend on the processor, memory, operating system, Python environment, library versions, and linear-algebra thread settings.

The cache-first policy was evaluated by repeating previously saved simulator and hardware tasks. Before execution, the agent constructs a task-dependent signature from the molecule, basis, ansätz, backend, shot count, SQD settings, mitigation policy, and software environment. If a compatible artifact exists, the agent summarizes the stored result rather than recomputing the workflow. For hardware tasks, saved job identifiers allow completed jobs to be retrieved after interruptions. This is particularly important for QPU experiments, where resubmission can consume additional queue time, shots, and backend access.

Convergence reporting is generated from the per-iteration SQD traces saved by the execution layer. The reporting agent can use these traces to produce energy-versus-iteration plots, subspace-size summaries, and ansätz-comparison tables. This supports the abstract-level claim that the framework provides convergence visualization and interactive reporting from stored artifacts rather than requiring the user to manually inspect raw output files.

\subsection{Hardware Execution and Mitigation Comparison}

To validate execution on a physical quantum device, the N\textsubscript{2}/LUCJ workflow
was executed on the 156-qubit IBM Quantum backend \texttt{ibm\_kingston} using
\num{100000} shots per job. The experiment used the full-space LUCJ ansätz, which corresponds to a 20-qubit logical circuit. The circuit was transpiled using a
VF2 calibration-aware layout with \texttt{opt\_level}=3, resulting in a
hardware circuit on 156 physical qubits with post-transpilation total depth \num{210} (which includes both single qubit gates and 2-qubit gates), the actual 2-qubit gate depth is \num{42}. We ran for multiple seeds and then pick the lowest depth quantum circuit for hardware execution. Also the post-transpilation 2qubit gates = 298. The hardware runner submitted one IBM Runtime job
per supported mitigation setting, retrieved the measurement results, applied the
same SQD post-processing used in simulator runs, and generated the mitigation
comparison reported in Table~\ref{tab:n2_hw_mitigation}. The saved artifacts
include the CSV/XLSX comparison table, metadata JSON, pre- and post-transpilation
circuit statistics, strategy-wise recovered energies, observed job-wall-time plus
SQD-diagonalization time, and job identifiers for retrieving interrupted or
previously submitted hardware jobs. Here, cost refers to the number of QPU jobs,
shot count, QPU runtime, retrieval overhead, and classical post-processing time.

\begin{table}[t]
\centering
\caption{Hardware error-mitigation comparison for N\textsubscript{2} with LUCJ ansätz on \texttt{ibm\_kingston}. Energies are compared against FCI as a reference. QPU+Diag (s) is the total time or observed QPU-job execution/wall time(including queue time on quantum device, actual quantum time plus classical SQD diagonalization time, in seconds. Note: only first job had queue time of 142 sec included, which is not there in any other jobs.}
\label{tab:n2_hw_mitigation}
\scriptsize
\renewcommand{\arraystretch}{1.12}
\setlength{\extrarowheight}{4pt}
\begin{tabular}{| c c c c |}
\hline
\textbf{Mitigation} & \textbf{$E_{\mathrm{SQD}}$ (Ha)} &
\textbf{$\lvert\Delta\rvert$ vs FCI (mHa)} & \textbf{QPU$+$Diag (s)} \\
\hline
None &
$-107.654103$ &
$0.01947697$ &
$149.4050$~s \\


DD &
$-107.6541206$ &
$0.001803794$ &
$48.606$~s \\

Twirling &
$-107.6539922$ &
$0.130246042$ &
$49.4789$~s \\

Twirling + DD &
$-107.6540009$ &
$0.121576141$ &
$50.1549$~s \\
\hline
\end{tabular}
\end{table}

For this molecule, backend, ansätz, and shot count, all tested mitigation
settings recover the FCI reference to within $10^{-3}$~mHa, while the
unmitigated baseline has the lowest observed runtime. We do not interpret this
as evidence that unmitigated execution is generally optimal, nor that the same
mitigation ranking will hold for larger molecules, deeper circuits, or different
backend calibration states. Instead, the result validates the framework's ability
to produce a context-dependent mitigation report that makes the observed
accuracy--cost trade-off explicit.


The hardware experiments and its results not only confirm to the simulator results or the state-of-art results one can generate by directly running SQD, but also support the main systems claims of SQD Agent: natural-language requests can be translated into SQD-family workflows; the same execution layer supports simulator and hardware paths; resource metrics and convergence traces are saved by default; cached artifacts can be reused; and mitigation strategies can be suggested or compared through hardware-specific reports. 

 The simulator benchmarks use small molecules in a minimal basis where FCI or CASCI references remain available. The hardware experiment uses one molecule, one ansätz, and one backend, namely N\textsubscript{2}/LUCJ on \texttt{ibm_kingston}. Therefore, the hardware results should be read as validation of the agentic capabilities,including hardware execution rather than as a broad chemistry benchmark. Profiling results are also machine-dependent and are reported as lightweight timing summaries rather than a complete low-level performance model. Finally, the agent is autonomous in workflow orchestration, but safety-critical choices such as fetched geometries, hardware submission, shot count, and mitigation strategy remain human-confirmed by design.

%% file: conclusion/merged.tex
\section{Discussions and Future Scope}
\label{sec:Conclusion}
In this work, we present an LLM-powered multi-agent system that dynamically generates and executes quantum chemistry workflows, thereby increasing the accessibility of quantum chemistry experiments. In the current work, we focused on the electronic structure prediction problem using the SQD algorithm. The agent can understand natural-language text instructions, generate the workflows accordingly, and, as per the user's request, run on a simulator or on real hardware. 
In hardware runs, it
can additionally set up and compare error-mitigation strategies and report the
associated accuracy and runtime costs. We evaluated it on $N_2$ 
and $CH_2O$
against CASCI and CCSD references on simulator and validated execution of $N_2$ on a 156-qubit IBM~Quantum processor, where the SQD energy reproduced the exact (FCI) result to well within chemical accuracy as shown in detail. We emphasize that the present work does not claim quantum advantage or introduce a new SQD algorithm; rather, its contribution is an orchestration framework that integrates established SQD-family algorithms, chemistry tools, classical post-processing, and quantum hardware into reproducible quantum--classical workflow. 

Beyond a single application, SQD Agent is built from modular Python components (\texttt{runner.py}, \texttt{chemistry.py}, \texttt{active\_space.py}) that extend to new molecular systems through additional molecular data and workflow specifications. Reproducibility is supported by a cache-first execution policy and timestamped artifacts containing commands, metadata, traces, metrics, and outputs.

The framework builds on the broader
move toward LLM-driven scientific agents, 
and carries it from the coordination of classical chemistry software into the quantum-native setting,
where the agent manages circuit execution and error mitigation on real devices as well as on
simulators. Its modular design is intended to scale from the small molecules studied here toward
larger, chemically richer targets such as transition-metal clusters ($[Fe_2S_2]$, $[Fe_4S_4]$),
catalytic centers, and biomolecular systems, which is the next step we are working towards. Such scaling reflects the portability of the orchestration framework and should not be
interpreted as a claim that SQD Agent changes the computational scaling or accuracy properties of the underlying SQD algorithms. We expect agent-based
frameworks of this kind to play a growing role in quantum-classical co-design, hardware-aware
algorithm selection, and adaptive sampling on near-term quantum devices.

The present study is bounded by available computational resources, so both the simulator and hardware experiments detailed in results section are restricted to small molecules in a minimal (STO-3G) basis, where full-space FCI remains computable as a reference. The hardware demonstration is a single molecule (N$_2$/LUCJ) on one backend (\texttt{ibm\_kingston}) and is included to validate
the execution path. Reported energies come from single runs with an unseeded sampler, so we quote no run-to-run
error bars, and reproducibility is provided through cached artifacts rather than bitwise-identical recomputation. Finally, the agent layer is an orchestration and
policy framework rather than an autonomous reasoner. Its reliability depends on the underlying LLM, and safety-critical steps such as hardware submission and the use of
fetched geometries are gated by explicit human confirmation by design. As future work, we plan to strengthen guardrails around LLM-based decision making to retain flexibility and creativity while reducing hallucinations and unsafe or scientifically inconsistent actions. The current framework records convergence traces, resource and timing information, profiling outputs, and mitigation reports that can inform subsequent workflow changes; however, reconfiguration is presently human-supervised rather than operating as an automatic closed-loop feedback system. We plan to develop such a closed-loop mechanism in which validated execution feedback can guide subsequent choices of ansatz, sampling parameters, backend configuration, and error-mitigation strategies. 
We are working on features such as adaptive error mitigation tailored to execution on physical quantum hardware, and a recommendation of which error mitigation schemes can be applied, what could be the range of typical error recovery with them, and the budget (compute time) they would need so that users can understand what algorithms they can explore while using this framework.
Agent efficiency will also be improved through token-optimization techniques. For example, a representative current LiH workflow required seven LLM calls and five tool calls, corresponding to 67,978 processed tokens; future versions will investigate more compact context management, selective retrieval, and ML-based methods for token-efficient planning and error-mitigation recommendation. Finally, the present work should be viewed as a proof-of-concept systems evaluation: preliminary use with a small expert user set was encouraging, but is insufficient for statistically meaningful conclusions about workflow-efficiency gains. We therefore plan a dedicated pilot user study measuring task-completion time, intervention frequency, success rate, usability, and workflow smoothness relative to direct/manual SQD execution. The current framework should be treated as a first version of multi-agent Quantum CoScientist and currently has only one algorithm family (SQD) for only one task molecular ground state estimation, but will be extended to several tasks with more algorithmic support as well as algorithm discovery in subsequent work.

%% file: Appendix/Appendix_v2.tex
\appendices
\section{Additional Benchmark Results}
\label{appendix:results}

This appendix reports supplementary SQD Agent benchmarks, including energies, runtimes, and circuit-size metrics for the molecular systems used in the evaluation. Energies are given in Hartree (Ha), deviations are reported relative to the full-space FCI reference in millihartree (mHa), and runtimes are given in seconds. For SQD results, ``full'' denotes the full orbital space, while ``active'' denotes the selected active-space window. Reported circuit metrics include qubit count, pre- and post-transpilation depth, two-qubit gate count, and total gate count; a dash indicates that the quantity is either not applicable or was not recorded. 
\begin{table}[h]
\centering
\caption{H\textsubscript{2}O benchmark results using full-space and active-space SQD variants.
Active-space window: $n_{\mathrm{core}}=2$, $n_{\mathrm{cas}}=5$, $n_{\mathrm{elecas}}=(3,3)$.}
\label{tab:h2o_energy_time}
\scriptsize
\resizebox{\columnwidth}{!}{%
\begin{tabular}{lccccc}
\toprule
\textbf{Method} & \textbf{Energy (Ha)} & \textbf{$\Delta$ vs FCI (mHa)} &
\textbf{Runtime (s)} & \textbf{\#Qubits} & \textbf{2Q (post)} \\
\midrule
RHF                 & -74.96294666 & +49.491 & 0.237 & -- & -- \\
MP2                 & -74.99844949 & +13.988 & 1.702 & -- & -- \\
CCSD                & -75.01232107 & +0.116  & 0.127 & -- & -- \\
CASCI (full)        & -75.01243743 & +0.000  & 0.085 & -- & -- \\
FCI (full)          & -75.01243743 & +0.000  & ---   & -- & -- \\
SQD (full) [UCJ]    & -75.01229993 & +0.138  & 0.966 & 14 & 3536 \\
SQD (active) [UCJ]  & -74.99663647 & +15.801 & 0.416 & 10 & 1096 \\
SQD (full) [LUCJ]   & -75.01229944 & +0.138  & 0.846 & 14 & 3516 \\
SQD (active) [LUCJ] & -74.99667086 & +15.767 & 0.405 & 10 & 1096 \\
SQD (full) [HE]     & -75.01243743 & +0.000  & 4.627 & 14 & 28 \\
SQD (active) [HE]   & -74.99667929 & +15.758 & 0.571 & 10 & 20 \\
SQD (full) [HF]     & -74.96294666 & +49.491 & 0.665 & 14 & 0 \\
SQD (active) [HF]   & -74.96294666 & +49.491 & 0.443 & 10 & 0 \\
CASCI (active)      & -74.99667929 & +15.758 & 0.013 & -- & -- \\
\bottomrule
\end{tabular}
}
\end{table}

\onecolumn

\begin{table}[!htbp]
\centering
\vspace{-1mm}
\caption{LiH benchmark results using full-space and active-space SQD variants.
For this molecule, the selected active space spans the full six-orbital space.}
\label{tab:lih_energy_time}
\scriptsize
\resizebox{\textwidth}{!}{%
\begin{tabular}{lccccccccccc}
\toprule
\textbf{Method} & \textbf{Energy (Ha)} & \textbf{$\Delta$ vs FCI (mHa)} & \textbf{Runtime (s)} &
\textbf{\#Qubits} & \textbf{Depth (pre)} & \textbf{Depth (post)} &
\textbf{2Q (pre)} & \textbf{2Q (post)} & \textbf{Total (pre)} & \textbf{Total (post)} \\
\midrule
RHF                 & -7.86186477 & +20.460 & 0.210 & -- & -- & -- & -- & -- & -- & -- \\
MP2                 & -7.87476887 & +7.556  & 1.233 & -- & -- & -- & -- & -- & -- & -- \\
CCSD                & -7.88231382 & +0.011  & 0.149 & -- & -- & -- & -- & -- & -- & -- \\
CASCI (full)        & -7.88232438 & +0.000  & 0.056 & -- & -- & -- & -- & -- & -- & -- \\
FCI (full)          & -7.88232438 & +0.000  & ---   & -- & -- & -- & -- & -- & -- & -- \\
SQD (full) [UCJ]    & -7.88229363 & +0.031  & 0.560 & 12 & 3  & 646 & 0  & 2012 & 2  & 3538 \\
SQD (active) [UCJ]  & -7.88224839 & +0.076  & 0.485 & 12 & 3  & 646 & 0  & 2012 & 2  & 3538 \\
SQD (full) [LUCJ]   & -7.88229363 & +0.031  & 0.487 & 12 & 3  & 578 & 0  & 1856 & 2  & 3354 \\
SQD (active) [LUCJ] & -7.88230837 & +0.016  & 0.469 & 12 & 3  & 578 & 0  & 1856 & 2  & 3354 \\
SQD (full) [HE]     & -7.88232438 & +0.000  & 1.250 & 12 & 28 & 27  & 24 & 24   & 49 & 48 \\
SQD (active) [HE]   & -7.88232438 & +0.000  & 1.510 & 12 & 28 & 27  & 24 & 24   & 49 & 48 \\
SQD (full) [HF]     & -7.86186477 & +20.460 & 2.186 & 12 & 2  & 2   & 0  & 0    & 1  & 4 \\
SQD (active) [HF]   & -7.86186477 & +20.460 & 0.648 & 12 & 2  & 2   & 0  & 0    & 1  & 4 \\
CASCI (active)      & -7.88232438 & +0.000  & 0.035 & -- & -- & -- & -- & -- & -- & -- \\
\bottomrule
\end{tabular}
}
\vspace{-3mm}
\end{table}

\begin{table}[!htbp]
\centering
\caption{N\textsubscript{2} benchmark results using full-space and active-space SQD variants.
Active-space window: $n_{\mathrm{core}}=4$, $n_{\mathrm{cas}}=6$, $n_{\mathrm{elecas}}=(3,3)$.}
\label{tab:n2_energy_time}
\scriptsize
\resizebox{\textwidth}{!}{%
\begin{tabular}{lccccccccccc}

\toprule

\textbf{Method} & \textbf{Energy (Ha)} & \textbf{$\Delta$ vs FCI (mHa)} & \textbf{Runtime (s)} &
\textbf{\#Qubits} & \textbf{Depth (pre)} & \textbf{Depth (post)} &
\textbf{2Q (pre)} & \textbf{2Q (post)} & \textbf{Total (pre)} & \textbf{Total (post)} \\

\midrule
RHF                 & -107.49650051 & +157.622 & 0.220 & -- & -- & -- & -- & -- & -- & -- \\
MP2                 & -107.65142137 & +2.701   & 2.526 & -- & -- & -- & -- & -- & -- & -- \\
CCSD                & -107.65019741 & +3.925   & 0.185 & -- & -- & -- & -- & -- & -- & -- \\
CASCI (full)        & -107.65412245 & +0.000   & 0.229 & -- & -- & -- & -- & -- & -- & -- \\
FCI (full)          & -107.65412245 & +0.000   & ---   & -- & -- & -- & -- & -- & -- & -- \\
SQD (full) [UCJ]    & -107.65326614 & +0.856   & 28.074 & 20 & 3 & 2788 & 0 & 15240 & 2 & 27524 \\
SQD (active) [UCJ]  & -107.62308076 & +31.042  & 0.522 & 12 & 3 & 818  & 0 & 2584  & 2 & 4176 \\
SQD (full) [LUCJ]   & -107.65218217 & +1.940   & 25.955 & 20 & 3 & 2695 & 0 & 14712 & 2 & 26468 \\
SQD (active) [LUCJ] & -107.62291478 & +31.208  & 0.575 & 12 & 3 & 816  & 0 & 2648  & 2 & 4274 \\
SQD (full) [HE]     & -107.64268871 & +11.434  & 17.601 & 20 & 44 & 43  & 40 & 40 & 81 & 80 \\
SQD (active) [HE]   & -107.62310177 & +31.021  & 1.329 & 12 & 28 & 27  & 24 & 24 & 49 & 48 \\
SQD (full) [HF]     & -107.49650051 & +157.622 & 1.105 & 20 & 2 & 2  & 0 & 0 & 1 & 14 \\
SQD (active) [HF]   & -107.49650051 & +157.622 & 0.658 & 12 & 2 & 2  & 0 & 0 & 1 & 6 \\
CASCI (active)      & -107.62310177 & +31.021  & 0.148 & -- & -- & -- & -- & -- & -- & -- \\
\bottomrule
\end{tabular}
}
\vspace{-3mm}
\end{table}

\begin{table}[!htbp]
\centering
\caption{CH\textsubscript{2}O benchmark results using the SQD Agent.}
\label{tab:h2co_energy_time}
\scriptsize
\resizebox{\textwidth}{!}{%
\begin{tabular}{lccccccccccc}
\toprule
\textbf{Method} & \textbf{Energy (Ha)} & \textbf{$\Delta$ vs FCI (mHa)} & \textbf{Runtime (s)} &
\textbf{\#Qubits} & \textbf{Depth (pre)} & \textbf{Depth (post)} &
\textbf{2Q (pre)} & \textbf{2Q (post)} & \textbf{Total (pre)} & \textbf{Total (post)} \\
\midrule
RHF                 & -112.34795500 & +139.017 & 0.193 & -- & -- & -- & -- & -- & -- & -- \\
MP2                 & -112.45997198 & +27.000  & 1.506 & -- & -- & -- & -- & -- & -- & -- \\
CCSD                & -112.48475257 & +2.219   & 0.255 & -- & -- & -- & -- & -- & -- & -- \\
CASCI (full)        & -112.48697170 & +0.000   & 2.075 & -- & -- & -- & -- & -- & -- & -- \\
FCI (full)          & -112.48697170 & +0.000   & ---   & -- & -- & -- & -- & -- & -- & -- \\
SQD (full) [UCJ]    & -112.48580783 & +1.164   & 1128.770 & 24 & 3 & 5444 & 0 & 35280 & 2 & 63304 \\

SQD (full) [LUCJ]   & -112.48308521 & +3.886   & 1125.513 & 24 & 3 & 5040 & 0 & 33612 & 2 & 61786 \\

SQD (full) [HE]     & -112.47995370 & +7.018   & 74.460 & 24 & 52 & 51 & 48 & 48 & 97 & 96 \\

SQD (full) [HF]     & -112.34795500 & +139.017 & 1.775 & 24 & 2 & 2 & 0 & 0 & 1 & 16 \\
SQD (active) [HF]   & -112.34795500 & +139.017 & 1.095 & 12 & 2 & 2 & 0 & 0 & 1 & 6 \\
\bottomrule
\end{tabular}
}
\vspace{-3mm}
\end{table}